\documentclass[12pt]{iopart}

\usepackage{graphicx}
\usepackage{dcolumn}
\usepackage{bm}
\usepackage{pifont}
\usepackage{color}

\usepackage{iopams}
\usepackage{amssymb,bm}

\usepackage{units}

\begin{document}

\title[Method for the determination of SOI constants]{Alternative method for the quantitative determination of Rashba- and Dresselhaus spin-orbit interaction using the magnetization}

\author{M~A Wilde and D Grundler}
\address{Lehrstuhl f\"ur Physik funktionaler Schichtsysteme,
Technische Universit\"at M\"unchen, Physik Department,
James-Franck-Str. 1, D-85747 Garching b. M\"unchen, Germany}

\ead{mwilde@ph.tum.de}

\begin{abstract}
The quantum oscillatory magnetization $M$ of a two-dimensional electron system in a magnetic field $\bi{B}$ is found to provide quantitative information on both the Rashba- and Dresselhaus spin-orbit interaction (SOI). This is shown by first numerically solving the model Hamiltonian including the linear Rashba- and Dresselhaus SOI and the Zeeman term in an in particular doubly tilted magnetic field and second evaluating the intrinsically anisotropic magnetization for different directions of the in-plane magnetic field component. The amplitude of specific magnetic quantum oscillations in $M(\bi{B})$ is found to be a direct measure of the SOI strength at fields $B$ where SOI-induced Landau level anticrossings occur. The anisotropic $M$ allows one to quantify the magnitude of both contributions as well as their relative sign. The influence of cubic Dresselhaus SOI on the results is discussed. We use realistic sample parameters and show that recently reported experimental techniques provide a sensitivity which allows for the detection of the predicted phenomena.
\end{abstract}

\pacs{73.21.Fg, 75,70.Tj, 85.75.-d}
\noindent{\it Keywords}: Spin-Orbit Interaction, Rashba effect, Dresselhaus effect, de Haas-van Alphen effect, Magnetization

\submitto{\NJP}
\maketitle
\tableofcontents

\section{Introduction}

The spin-orbit interaction (SOI) in low-dimensional electron systems in semiconductors has been investigated intensely both theoretically and experimentally due to the rich spin physics and potential application in the field of spintronics \cite{Wolf2001,Grundler2002,Zutic2004}. In a two-dimensional electron system (2DES) spin-orbit interaction arises due to either structure inversion asymmetry or bulk inversion asymmetry of the host crystal. These contributions are known as Rashba (R) and Dresselhaus (D) SOI, respectively. Early spintronics proposals \cite{Datta1990} focused on the R-SOI that can be tuned by an external electric field \cite{Nitta1997}. Here, the envisaged devices were based on ballistic transport. More recently, proposals emerged that rely on the interplay of R-SOI and D-SOI \cite{Schliemann2003,Cartoix`a2003} and work for diffusive transport. In any case, a prerequisite for the realization of spintronic devices is the unambiguous and quantitative knowledge of R-SOI and D-SOI present in a given electron system. This evaluation is still challenging. A widespread method for the determination of SOI coupling constants is based on the oscillatory resistivity $\rho(B)$ measured in a magnetic field $\bi{B}$ at low temperature $T$. The analysis of the relevant beating patterns does not yield unambiguous results however, when both R-SOI and D-SOI play a comparable role. Further on, the weak-antilocalization phenomenon in $\rho$ near $B=0$ is used to determine SOI coupling constants. Here, different models have been put forward which help to evaluate experimental data only in specific parameter regimes \cite{Hikami1980,Kawabata1984,Iordanskii1994,Knap1996,Miller2003,Golub2005,Kallaher2010,Faniel2011}. Spin photocurrents \cite{Ganichev2004} provided the ratio and relative sign, but the absolute values of the SOI coupling constants were not accessible. Recently, the R-SOI and D-SOI strength was inferred from spin precessional motion as a function of an in-plane electric field monitored by time-resolved Faraday rotation \cite{Meier2007} and from simultaneous measurements of the in-plane electron $g$-factor and the spin relaxation rate by spin-quantum beat spectroscopy \cite{Eldridge2011}. In a pioneering theoretical work \cite{Fal'ko1992} Fal'ko considered cyclotron and electric-dipole spin resonance in tilted magnetic fields and provided an approach to extract both SOI coupling constants. The author developed an analytical expression for the anticrossing energy gap depending on both the R-SOI and D-SOI contributions but left the exact diagonalization of the problem outside the scope of the paper.

Magnetization experiments addressing in particular the ground state of a 2DES have so far not been thoroughly discussed in the framework of spintronics and the separate extraction of R-SOI and D-SOI coupling constants. This is surprising as magnetization measurements have been used already to gain quantitative insight into 2DES properties such as, e.g., quantum lifetimes \cite{Schwarz2002}, self-consistent electron redistribution \cite{Schaapman2003}, giant spin splitting \cite{Harris2001} and electron-electron interaction \cite{Bominaar-Silkens2006}. Early on, theoretical works predicted beating patterns in the quantum oscillations of $M$ due to SOI \cite{Bychkov1984,eSilva1994,Zawadzki2002,Wang2009} but corresponding magnetization measurements have been reported only recently \cite{Rupprecht2013,Wilde2009}. A rigorous theoretical treatment of $M(\bi{B})$ and its in-plane anisotropy resulting from the two SOI contributions in tilted magnetic fields has not yet been presented. The relevance of $M(\bi{B})$ for the quantitative determination of both SOI coupling constants has not been profoundly explored at all. In contrast to other techniques magnetization experiments do not address excited states and no knowledge about microscopic details of the scattering processes in the sample is needed. A detailed theoretical analysis and description of $M(\bi{B})$ is now timely as the magnetization is a non-invasive probe of electronic band structure properties that does not require contacts to the sample on the one hand and measures the system in equilibrium on the other hand.

In this paper we address the oscillatory magnetization $M(\bi{B})$ of a 2DES theoretically in detail and show that in particular tilted field experiments can be used to extract both, R-SOI and D-SOI coupling constants as well as their relative sign. For this we perform numerical and analytical calculations based on the Hamiltonian first written out by Das \emph{et al.} \cite{Das1990}. We focus on the numerical diagonalization of the Hamiltonian including linear R-SOI, D-SOI terms and the Zeeman term in an arbitrarily tilted magnetic field. We further follow the analytical approaches of \cite{Fal'ko1992,Olendski2008}. In magnetic fields collinear with the 2DES normal, Landau levels (LLs) with different spin indices exhibit multiple crossings when SOI is present. The uneven spacing of LLs at the Fermi energy leads to the well-known beating patterns in the quantum oscillations. When the field direction is tilted away from the 2DES normal an anticrossing of LLs occurs \cite{Bychkov1990} that is caused by an SOI-induced mixing of levels. The anticrossing opens an energy gap in the ground-state energy spectrum when the Zeeman splitting equals the cyclotron energy. We will show that the gap provokes a characteristic quantum oscillation of $M$ whose amplitude $\Delta M$ is a direct measure of the SOI coupling constant if one SOI term dominates. For the case where both R-SOI and D-SOI terms play a role at the same time, the spectrum of $\Delta M$ is shown to be anisotropic with respect to the in-plane magnetic field direction. From the amplitude $\Delta M$ and its in-plane anisotropy, both, the R-SOI and D-SOI coupling constants are extracted quantitatively and their relative sign is determined. We treat the problem with exact diagonalization and present analytical approximations where possible to provide equations for the evaluation of experimental data. The full numerical treatment is shown to be in particular important for the detailed comparison with measured traces $M(\bi{B})$ in order to unambiguously identify the SOI contributions. As the magnetization $M$ is calculated directly from a model Hamiltonian without further assumptions, it is thus ideally suited for a quantitative modeling of experimental data \cite{Wilde2010}.

The paper is organized in the following way: In section \ref{sec:energyspectrum} we define the model Hamiltonian of the problem and calculate the quantum oscillatory magnetization with emphasis on the anticrossing behavior in the regime of high magnetic fields and high tilt angles. The results are put into relation to the well-known SOI-induced beating patterns occurring in quantum oscillations in the low-field regime. In section \ref{sec:application} the model calculations are applied to specific relevant systems. We discuss the feasibility of the proposed experiments and the modeling of experimental data in section \ref{sec:discussion}. Finally, we draw conclusions in section \ref{sec:conclusions}.

\section{Energy spectrum and magnetization of 2DESs with R-SOI and D-SOI in tilted magnetic fields}\label{sec:energyspectrum}

The magnetization $M=-\partial F/\partial B|_{T,n_{\textrm{s}}}$ is a thermodynamic state function that can directly be linked to a model Hamiltonian $H$. The function reduces to $M=-\partial U/\partial B|_{T=0,n_{\textrm{s}}}$ at $T=0$~K. Here $F$ ($U$) denotes the free (ground state) energy of the system and $n_{\textrm{s}}$ is the sheet electron density. In the following we formulate the problem for the exact numerical calculation of the quantum oscillatory magnetization $M$, i.e., the de Haas-van Alphen (dHvA) effect, and give analytical approximations that allow for detailed simulations of experimental data.

\subsection{Model Hamiltonian}

In the following we consider zinc blende semiconductor quantum wells grown in the $[001]$ direction. This class of 2DESs is most relevant for conducting experiments. The calculations are based on the model Hamiltonian $H=H_0+H_{\textrm{SO}}$, taking into account the linear Rashba (R-SOI) term and the linear Dresselhaus (D-SOI) term. The latter arises from the $k^3$ bulk spin-orbit interaction and is the relevant term in the limit of narrow quantum wells and small in-plane wave vectors $k$ of the electrons in the conduction band. Additional effects due to the $k^3$ D-SOI relevant in the limit of wide quantum wells and large $k$ will be discussed in section~\ref{sec:cubicDSOI}. The Hamiltonian for the $k$-linear terms reads:
\begin{equation}
H=H_0+\frac{\alpha_{\textrm{R}}}{\hbar} \left( \sigma_x \pi_y - \sigma_y \pi_x \right)+ \frac{\beta_{\textrm{D}}}{\hbar} \left( \sigma_x \pi_x - \sigma_y \pi_y \right) \mbox{ .} \label{eq:hamiltonian}
\end{equation}
Here $\pi_{x,y}=p_{x,y}+eA_{x,y}$ where $p_{x,y}=-\imath \hbar \case{\partial}{\partial (x,y)}$ are the components of the in-plane momentum operator, $A_{x,y}$ are the components of the vector potential, $e$ is the electron charge and $\alpha_{\textrm{R}}$ ($\beta_{\textrm{D}}$) is the R-SOI (D-SOI) spin-orbit coupling strength. $H_0=\frac{\bm{\pi}^2}{2m^*}+\frac{1}{2}g^*\mu_B \bi{B} \bm{\sigma}$, $\bm{\pi}=(\pi_x,\pi_y)$ includes the Zeeman contribution with the effective Land\'{e} factor $g^*$, the Bohr magneton $\mu_{\textrm{B}}=e\hbar/2m_e$ and the vector of Pauli spin matrices $\bm{\sigma}=(\sigma_x,\sigma_y,\sigma_z)$. Here, $m^*$ is the effective mass, and $m_e$ is the free electron mass. The effects of $g$-factor anisotropy will be discussed in section~\ref{sec:discussion} The magnetic field is chosen as $\bi{B}=B\bi{b}=B(\sin \theta \cos \phi, \sin \theta \sin \phi, \cos \theta)$, where $\theta$ is the tilt angle between $\bi{B}$ and the 2DES normal $\bi{n}$ and $\phi$ is the azimuthal angle with respect to the $[100]$ direction. We consider the lowest subband of the narrow quantum well to be occupied.

In the presence of both SOI terms in tilted magnetic fields no analytical solution is known. In this paper, we focus on the exact numerical diagonalization of $H$. For this, we use the matrix elements of $H$ between the eigenstates of $H_0$ in perpendicular magnetic fields, i.e., between the well known Landau states $|n,\sigma \rangle$ with LL index $n=0,1,2,\ldots$ and energies $E_n^{\sigma}=(n+\nicefrac{1}{2})\hbar \omega_{\textrm{c}}+\frac{1}{2}g^*\mu_{\textrm{B}}B\sigma$. Here, $\omega_{\textrm{c}}=eB_{\perp}/m^*$ is the cyclotron frequency and $B_{\perp}$ is the magnetic field component perpendicular to the 2DES. The corresponding matrix elements have been first written out by Das \emph{et al.} \cite{Das1990}.\footnote{Note that for $\alpha_{\textrm{R}} \neq 0,\beta_{\textrm{D}} \neq 0,g^* \neq 0,\theta \neq 0$, the energy spectrum is anisotropic with respect to the direction of the in-plane magnetic field component, and the Zeeman terms in $H_0$ have to take into account the in-plane field direction. This case has not been discussed in \cite{Das1990}.} The resulting matrix is diagonalized numerically by truncating the matrix dimensions while including a sufficient number of LLs. It is instructive, however, to reexpress the Hamiltonian (\ref{eq:hamiltonian}) in a basis where the spin quantization axis is chosen along the magnetic field direction following \cite{Fal'ko1992,Olendski2008}. While this bears no particular advantage for the exact numerical diagonalization, it allows us to gain additional analytical insight into the anticrossing as we will discuss below. After the rotation
\begin{eqnarray}
\sigma_x \rightarrow & \sigma_x \cos \theta \cos \phi - \sigma_y \sin \phi + \sigma_z \sin \theta \cos \phi \nonumber \\
\sigma_y \rightarrow & \sigma_x \cos \theta \sin \phi - \sigma_y \cos \phi + \sigma_z \sin \theta \sin \phi  \label{eq:rotation} \\
\sigma_z \rightarrow & -\sigma_x \sin \theta + \sigma_z \cos \theta \nonumber
\end{eqnarray}
(\ref{eq:rotation}) yields for the spin-orbit part of (\ref{eq:hamiltonian}) \cite{Olendski2008}
\begin{eqnarray}
H_{\textrm{SO}} = & \frac{\pi_+}{2} \left[ \sigma_+ \left( \eta_++\eta_-\cos \theta \right)-\sigma_- \left( \eta_+-\eta_- \cos \theta \right) \right. \nonumber \\
& + \left. \sigma_z \eta_- \sin \theta \right] + \mbox{H.c. ,} \label{eq:H-SO}
\end{eqnarray}
where we used $\sigma_{\pm}=(\sigma_x \pm \imath \sigma_y)/2$, $\pi_{\pm}=\pi_x \pm \imath \pi_y$ and the convenient notation $\eta_{\pm}=\beta_{\textrm{D}}e^{\imath \phi} \pm \imath \alpha_{\textrm{R}}e^{-\imath \phi}$ introduced in \cite{Olendski2008}. The corresponding nonvanishing matrix elements can be written as
\begin{eqnarray}
\langle n,\pm |H_{0}|n,\pm \rangle &= \hbar \omega_{\textrm{c}} (n+\frac{1}{2}) \pm \frac{1}{2} g^*\mu_{\textrm{B}}B \label{eq:el_diagonal} \\
\langle n+1,\pm|H_{\textrm{SO}}|n,\mp \rangle &= \pm \imath \sqrt{\frac{n+1}{2l_{\textrm{B}}^2}} \left[ \eta_+(\phi) \pm \eta_-(\phi) \cos \theta \right] \label{eq:el_olendski} \\
\langle n+1,\pm|H_{\textrm{SO}}|n,\pm \rangle &= \pm \imath \sqrt{\frac{n+1}{2l_{\textrm{B}}^2}} \eta_-(\phi) \sin \theta \mbox{ ,} \label{eq:el_2olendski}
\end{eqnarray}
and their Hermitian conjugates. Here, we have introduced the magnetic length $l_{\textrm{B}}=(\hbar/eB_{\perp})^{1/2}$.

\begin{figure}
\centering
\includegraphics{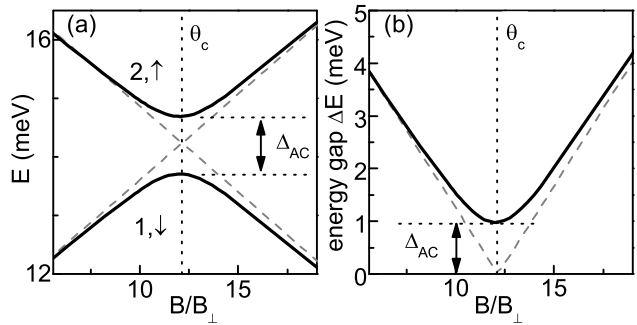}
\caption{\label{Fig1}(a) Anticrossing of levels (1,$\downarrow$), (2,$\uparrow$) (solid lines) numerically calculated for a $10$~nm wide InGaAs/InP quantum well. Sample parameters are $g^*=-4.45$, $m^*=0.037m_{\textrm{e}}$, $n_{\textrm{s}}=2.2\times 10^{15}$~m$^{-2}$, $\alpha_{\textrm{R}}=9.5\times 10^{-12}$~eVm and  $\beta_{\textrm{D}}=0$. For vanishing SOI the levels coincide (dashed lines) at an angle $\theta_{\textrm{c}}$. (b) Energy gap $\Delta E$ between levels in (a) as a function of $B/B_{\perp}=1/\cos \theta$. Note that the gap $\Delta_{\textrm{AC}}$ at $\theta_{\textrm{c}}$ is solely due to SOI.}
\end{figure}

\begin{figure}
\centering
\includegraphics[width=15cm]{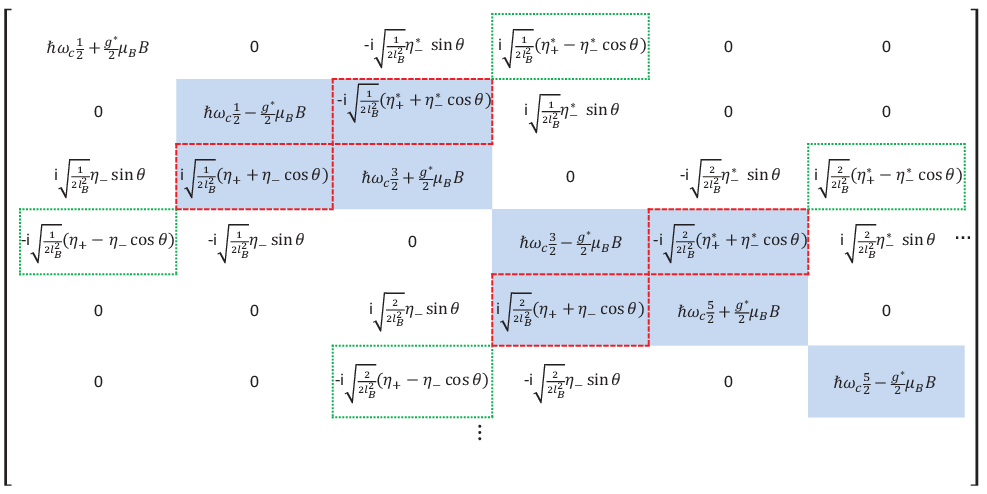}
\caption{\label{Fig2} Matrix of the problem. For a negative $g$-factor the coincidence condition in the absence of SOI is met when the diagonal matrix elements inside the shaded $2\times 2$ blocks are identical. When SOI is present, these diagonal matrix elements are coupled to first order by the matrix elements highlighted by dashed lines. Neglecting higher-order couplings, this leads to the analytical approximation for the anticrossing gap $\Delta_{\textrm{AC}}$ given by (\ref{eq:anticrossing-gap}). For a positive $g$-factor the corresponding diagonal terms are coupled by the matrix elements highlighted by dotted lines. }
\end{figure}

In a system without SOI the interplay of Landau quantization energy $(n+\nicefrac{1}{2})\hbar eB_{\perp}/m^*$ which depends on $B_{\perp}$ and Zeeman energy splitting $g^*\mu_{\textrm{B}} B$ which depends on the total $B=B_{\perp}/\cos \theta$ leads to artificial degeneracies between spin-split LLs whenever $l\cos \theta_{\textrm{c}} = (g^*/2)(m^*/m_{\textrm{e}})$ ($l=1,2,3\ldots$). Two levels with different Landau and spin indices ($\uparrow$, $\downarrow$) thus cross as a function of $B/B_{\perp}=1/\cos \theta$. A crossing between e.g. ($n=1$,$\downarrow$) and ($n=2$,$\uparrow$) LLs is sketched as dashed lines in figure~\ref{Fig1} (a) assuming InGaAs/InP quantum well (QW) parameters as reported by Guzenko \emph{et al.} \cite{Guzenko2007} who found that D-SOI is negligible in their samples. The condition for crossing \cite{Fang1968,Nicholas1988} - commonly named coincidence condition - has been used extensively in experiments to measure $|g^*|m^*$ \cite{Nicholas1988,Papadakis1999}.
 When SOI is present, an anticrossing gap $\Delta_{\textrm{AC}}$ is opened at the coincidence angle $\theta_{\textrm{c}}$ due to the resonant level mixing (solid lines). The gap $\Delta E$ between the levels (1,$\downarrow$) and (2,$\uparrow$) at integer filling factor $\nu=n_{\textrm{s}}/(eB/h)=4$ shown in figure~\ref{Fig1} has a finite value $\Delta_{\textrm{AC}}$ at $\theta_{\textrm{c}}$ [figure~\ref{Fig1} (b)] (whereas $\Delta E=0$ would be expected without SOI). At this angle $\theta_{\textrm{c}}$, the SOI-induced mixing of levels is strong. To first order, the two levels are coupled by the matrix elements as illustrated in figure~\ref{Fig2}.
 Following this, the anticrossing gap $\Delta_{\textrm{AC}}$ is approximately given by $2|\langle n+1,\pm|H_{\textrm{SO}}|n,\mp \rangle|$ at $\theta_{\textrm{c}}$, i.e., we have
\begin{eqnarray}
& \Delta_{\textrm{AC}} \approx \frac{\sqrt{2(n+1)}}{l_{\textrm{B}}} \nonumber \\
&  \times \sqrt{\beta_{\textrm{D}}^2 (1-\frac{g^*}{|g^*|}b_z )^2+4\alpha_{\textrm{R}} \beta_{\textrm{D}} b_xb_y+\alpha_{\textrm{R}}^2 (1+\frac{g^*}{|g^*|}b_z)^2 } \mbox{ .}\label{eq:anticrossing-gap}
\end{eqnarray}
This is consistent with the results of \cite{Fal'ko1992,Olendski2008}. $\Delta_{\textrm{AC}}$ is anisotropic with respect to the direction of the in-plane magnetic field component.

\subsection{Quantum oscillatory magnetization}

\begin{figure}
\centering
\includegraphics{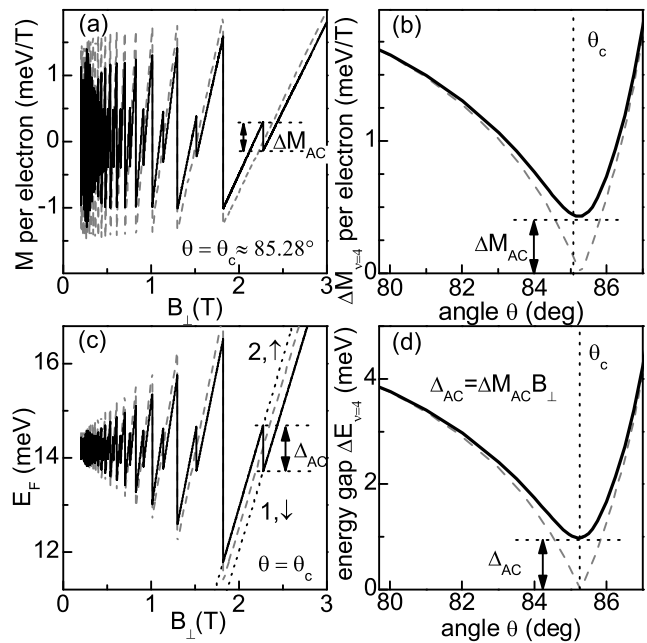}
\caption{\label{Fig3} (a) Magnetization $M$ numerically calculated at the coincidence angle $\theta_{\textrm{c}}$. Parameters as in figure~\ref{Fig1} (a) for $\alpha_{\textrm{R}}=9.5\times 10^{-12}$~eVm and $\beta_{\textrm{D}}=0$ (solid line). The peak-to-peak magnetization amplitude at $\nu=4$, $\Delta M_{\nu=4}(\theta_{\textrm{c}})=\Delta M_{\textrm{AC}}$, is marked by an arrow. (b) $\Delta M_{\nu=4}$ as a function of tilt angle $\theta$ (solid line). For vanishing SOI (dashed line) the amplitude $\Delta M$ reaches zero at the coincidence angle. For $\alpha_{\textrm{R}}\neq 0$ the minimal amplitude at $\theta_{\textrm{c}}$ (arrow) directly reflects the corresponding anticrossing gap $\Delta_{\textrm{AC}}=\Delta M_{\textrm{AC}}B_{\perp}$ highlighted in the oscillatory Fermi energy $E_{\textrm{F}}$ shown in (c). Dotted lines in (c) indicate the Landau levels $(2,\uparrow)$ and $(1,\downarrow)$. (d) $\Delta E_{\nu=4}$ as a function of tilt angle. Dashed lines are always for $\alpha_{\textrm{R}}=\beta_{\textrm{D}}=0$ for comparison.}
\end{figure}

In the following we relate $\Delta_{\textrm{AC}}$ to the magnetization $M$ as the experimentally accessible quantity. To calculate $M$, first the field-dependent Fermi energy $E_{\textrm{F}}$ is determined from the condition of constant density $n_{\textrm{s}}$ using the energy eigenvalues $\epsilon_n$. At $T=0$~K the ground state energy $U$ is calculated by summing up the energies of all occupied states. The magnetization perpendicular to the 2DES is obtained through $M=-\partial U/\partial B_{\perp}|_{n_{\textrm{s}},T=0}$ [figure~\ref{Fig3}~(a)] \footnote{The jump $\Delta M$ corresponding to an anticrossing gap $\Delta_{\textrm{AC}}$ occurs in the magnetization component collinear with the 2DES normal.}. The magnetization is found to oscillate with a sawtooth-like waveform that is characteristic of the dHvA effect in a 2DES \cite{Wilde2006}. To show how the anticrossing gap is reflected in the magnetization [figure~\ref{Fig3} (b)] we consider an InGaAs/InP sample with parameters as in figure~\ref{Fig1} at the coincidence angle $\theta_{\textrm{c}}$ ($\alpha_{\textrm{R}}=9.5\times 10^{-12}$~eVm).
 $M$ jumps discontinuously in figure~\ref{Fig3} (a) at integer filling factors $\nu$, whenever the Fermi energy $E_{\textrm{F}}$ jumps to the next lower lying LL with increasing magnetic field. The peak-to-peak amplitude $\Delta M$, normalized to the electron number $N=n_{\textrm{s}}\cdot A$, where $A$ is the area of the 2DES, is directly related to the jump $\Delta E$ in $E_{\textrm{F}}$ (i.e., the energy gap) via $\Delta M = \Delta E/B_{\perp}$ \cite{Wiegers1997,MacDonald1986,Manolescu1995}. Thus, the anticrossing manifests itself as a sharp jump in $M$, denoted by the arrow in figure~\ref{Fig3}~(a). $\Delta M$ exhibits the minimum amplitude $\Delta M_{\textrm{AC}}$ at $\theta=\theta_{\textrm{c}}$ [figure~\ref{Fig3}(b)] and is directly related to the anticrossing gap $\Delta_{\textrm{AC}}$ [figure~\ref{Fig3} (c) and (d)] via
\begin{equation}
\Delta M_{\textrm{AC}}=\Delta_{\textrm{AC}}/B_{\perp} \label{eq:maxwell} \mbox{ .}
\end{equation} In the case of finite SOI, the critical angle $\theta_{\textrm{c}}$ is found experimentally by looking for the minimum magnetization amplitude $\Delta M=\Delta M_{\textrm{AC}}$ in large $B$ providing the minimum gap size. Dashed lines in figure~\ref{Fig3} show the behaviour for $\alpha_{\textrm{R}}=\beta_{\textrm{D}}=0$. For $\theta=\theta_{\textrm{c}}$, oscillations occur only at odd integer $\nu$ [figure~\ref{Fig3} (a) and (c)] since the coincidence condition is met for even $\nu$ and an anticrossing gap does not form.

\subsection{Relation to SOI-induced beating patterns in quantum oscillations at small $B$}
\begin{figure}
\centering
\includegraphics{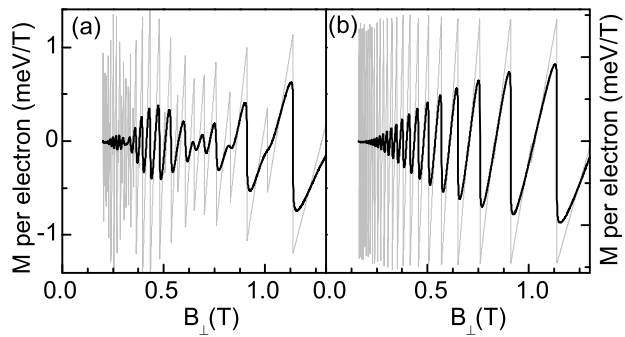}
\caption{\label{Fig4} (a) Beating pattern in $M$ for the InGaAs/InP QW containing an ideal 2DES considered above (light curve). The angle $\theta=15^{\circ}$ is fixed and $B_{\perp}$ is varied. The beating pattern is more clearly seen if a real 2DES is considered where electrons undergo scattering. To account for the scattering and LL broadening, a Gaussian level broadening of $\Gamma=0.3$~meV is assumed (dark curve). (b) $M$ for a 2DES with $\alpha_{\textrm{R}}=\beta_{\textrm{D}}=5\times 10^{-12}$~eVm. All other parameters are chosen like in (a). Here, the LL spectrum is linear and a beating pattern in $M$ does not occur although SOI is present.}
\end{figure}

Beating patterns in $M$ at small $B$ are a consequence of the nonlinear LL dispersion induced by the SOI, leading to small $\Delta M$ (nodes) when adjacent levels at $E_{\textrm{F}}$ are equally spaced, i.e., where the total spin splitting at the Fermi level - denoted by $\delta$ in the following - is $\delta =(p+ \nicefrac{1}{2}) \hbar \omega_{\textrm{c}}$, with $p=0,1,2,\ldots$ \cite{Zawadzki2004}. The beating pattern is illustrated in figure~\ref{Fig4} (a) considering the sample parameters of figure~\ref{Fig1}. Most earlier investigations have focused on the low field regime, and evaluated such beating patterns occurring in magneto-oscillations of $\rho$ \cite{Luo1990,Das1990,Nitta1997,Engels1997,Grundler2000}. However, as noted above, their analysis is sometimes ambiguous, especially when both $\alpha_{\textrm{R}}$ and $\beta_{\textrm{D}}$ play a role. To illustrate the ambiguity, we address the relevant case $\alpha_{\textrm{R}}=\beta_{\textrm{D}}$ (cf. Schliemann \emph{et al.} \cite{Schliemann2003} and Cartoix`a \emph{et al.} \cite{Cartoix`a2003}) in figure~\ref{Fig4} (b).
 The resulting LL spectrum is linear in $B$ as in the case $\alpha_{\textrm{R}}=\beta_{\textrm{D}}=0$, and a beating pattern is not formed. This way, no information about SOI can be gained in small fields $B$. Our theoretical approach and analysis outlined above predict however that for large $\bi{B}$, $\Delta M$ still experiences the anticrossing of LLs. By measuring the amplitude $\Delta M_{\textrm{AC}} (\phi)$ one still obtains information about SOI, even if the well-known beating is absent.

\section{Application to specific electron systems}\label{sec:application}

Of great interest in the field of spintronics are in particular two cases where (i) a strong (tunable) R-SOI dominates and D-SOI can be neglected \cite{Datta1990} and where (ii) R-SOI and D-SOI are of comparable or even identical strength \cite{Schliemann2003,Cartoix`a2003}. We will discuss the behaviour of $\Delta M$ for the two relevant scenarios in section~\ref{sec:Dominant_SOI_term} and section~\ref{sec:Comparable R-SOI and D-SOI}, respectively.

\subsection{2DES with R-SOI}\label{sec:Dominant_SOI_term}

\begin{figure}
\centering
\includegraphics{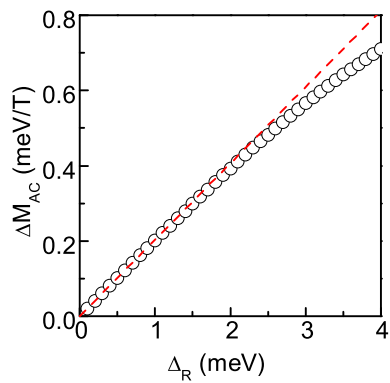}
\caption{\label{Fig5} Oscillation amplitude $\Delta M_{\textrm{AC}}$ at $\nu=4$ and $\theta=\theta_{\textrm{c}}$ vs Rashba splitting parameter  $\Delta_{\textrm{R}}$. Open symbols are the results of the exact diagonalization. The dashed line marks the first-order analytical approximation of equation (\ref{eq:delta-r-d}). The deviation from the exact result increases with increasing $\Delta_{\textrm{R}}$. Equation (\ref{eq:delta-r-d}) is a good approximation for $\Delta M_{\textrm{AC}}B_{\perp}/\hbar \omega_{\textrm{c}} \ll 1$.
}
\end{figure}

We first focus on the case where D-SOI can be neglected and R-SOI dominates. The result from the exact diagonalization is shown as open symbols in figure~\ref{Fig5}. The dependence deviates from a linear slope for $\Delta_{\textrm{R}}\geq 2$~meV. Since $\Delta M_{\textrm{AC}}$ is related to the energy gap $\Delta_{\textrm{AC}}$ at the Fermi energy $E_{\textrm{F}}$ we can reexpress (\ref{eq:anticrossing-gap}) in terms of the zero-field spin splitting energy at $E_{\textrm{F}}$ given by $\Delta_{\textrm{R}}=2\alpha_{\textrm{R}}k_{\textrm{F}}$. Here, $k_{\textrm{F}}=\sqrt{2\pi n_{\textrm{s}}}$ is the Fermi wave vector of the 2DES. Using (\ref{eq:maxwell}) the analytical approximation to the amplitude of the magnetic oscillation simplifies to
\begin{equation}
\Delta M_{\textrm{AC}} \approx \frac{\sqrt{\Delta_{\textrm{R}}^2 (1+\frac{g^*}{|g^*|}b_z )^2}}{B_{\perp}}  \mbox{ .}\label{eq:dM_one_dominant}
\end{equation}
In this approximation $\Delta M_{\textrm{AC}}$ is a linear measure of the SOI coupling constant that is independent of the direction of the in-plane magnetic field component. For the InGaAs/InP quantum well considered in figure~\ref{Fig1} $g^*$ is negative.\footnote{A negative $g^*$ can be assumed for systems with large SOI, see \cite{eSilva1994}.} In this case, we get
\begin{equation}
\Delta M_{\textrm{AC}} \approx \Delta_{\textrm{R}} \frac{\sin^2 \frac{\theta_{\textrm{c}}}{2}}{B_{\perp}} \mbox{ .} \label{eq:delta-r-d}
\end{equation}
This approximate analytical result for $\Delta M_{\textrm{AC}}$ is shown as a function of $\Delta_{\textrm{R}}=2\alpha_{\textrm{R}}k_{\textrm{F}}$ (dashed line) in figure~\ref{Fig5}.
 We evaluate the gap $\Delta_{\textrm{AC}}$ at the small filling factor $\nu=4$ here, since it occurs at sufficiently high $B_{\perp}$ for the total spin splitting $\delta$ to approach $\hbar \omega_{\textrm{c}}$ for the anticipated coincidence condition on the one hand and is still within experimental reach [cf. section.~\ref{sec:discussion}] on the other hand. By comparing the analytical with the exact result we find that (\ref{eq:delta-r-d}) is valid for $\Delta M_{\textrm{AC}}B_{\perp}/\hbar \omega_{\textrm{c}} \ll 1$. Higher order couplings become significant in the limit of large SOI coupling constants and small cyclotron energy $\hbar \omega_{\textrm{c}}$. These are not included in (\ref{eq:delta-r-d}) but relevant for the exact calculation. Equation~\ref{eq:delta-r-d} overestimates the real gap size. In the analysis of experimental data, (\ref{eq:dM_one_dominant}) can be used as a starting point for modeling of the magnetization. Analysis and modeling of the specific purely SOI-induced magnetic oscillation at, e.g., $\nu=4$ allows an independent determination of the Rashba constant that is complementary to the analysis of beating patterns in small magnetic fields. The analysis is applicable even in situations where beating patterns are not resolved. For $\Delta_{\textrm{R}}=0$ and $\Delta_{\textrm{D}}=2\beta_{\textrm{D}}k_{\textrm{F}}\neq 0$ one needs to replace $\Delta_{\textrm{R}}$ by $\Delta_{\textrm{D}}$ in this section and apply a minus sign before $\case{g^*}{|g^*|}$ in (\ref{eq:dM_one_dominant}).

\subsection{2DES with comparable R-SOI and D-SOI}\label{sec:Comparable R-SOI and D-SOI}

We now turn to the case where R-SOI and D-SOI are of the same order, but consider different relative strengths. For zero in-plane magnetic field, the magnetization for this case has been treated in \cite{eSilva1994}. Here, we focus on the general case of an arbitrarily tilted magnetic field applied under angles $\theta$ and $\phi$ defined above.

\begin{figure}
\centering
\includegraphics{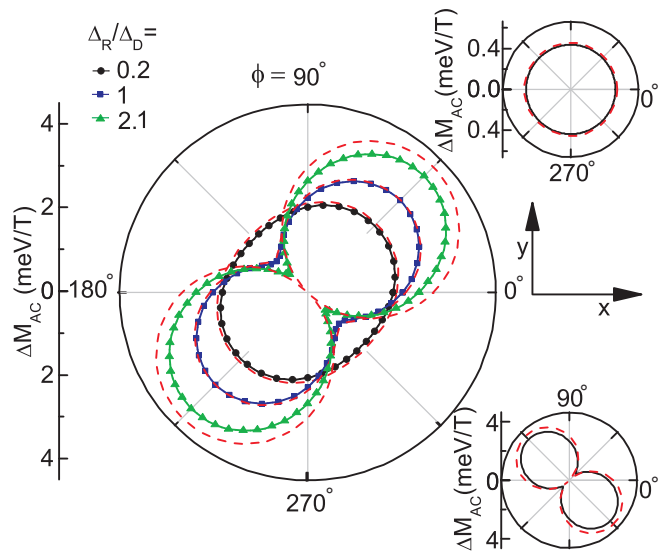}
\caption{\label{Fig6} Polar plot of $\Delta M_{\textrm{AC},\nu=4}$ vs azimuthal angle $\phi$ for an InSb QW for which we assume different R-SOI values (symbols interconnected by a solid line). InSb QW parameters are $m^*=0.014m_{\textrm{e}}$, $g^*=-51$, $n_{\textrm{s}}=2.5\times 10^{15}$~m$^{-2}$ and $\Delta_{\textrm{D}}=8.07$~meV. $\Delta_{\textrm{R}}$ takes the values $0.2, 1, \simeq 2.1 \times \Delta_{\textrm{D}}$ (symbols are defined in the top left inset). Note that for $\Delta_{\textrm{R}}/\Delta_{\textrm{D}}=1$ beating patterns do not exist in a perpendicular $B$. Still, $\Delta M_{\textrm{AC}}$ (squares) exhibits a characteristic anisotropy. Both, the size and relative sign of $\Delta_{\textrm{R}}$ and $\Delta_{\textrm{D}}$ determine the anticrossing gap value and anisotropy. For $\Delta_{\textrm{R}}/\Delta_{\textrm{D}}=-2.1$, $\Delta M_{\textrm{AC},\nu=14}$ vs $\phi$ is rotated by $\pi/2$ (bottom right inset) compared to $\Delta_{\textrm{R}}/\Delta_{\textrm{D}}=+2.1$. Top right inset: Result for the InGaAs/InP QW of figure \ref{Fig1}. $\Delta M_{\textrm{AC}}$ is isotropic since $\Delta_{\textrm{D}}=0$. The dashed lines in all plots correspond to the approximate solutions (\ref{eq:anisotropic-gap}). For large SOI splitting energies $(\Delta_{\textrm{R}}/\Delta_{\textrm{D}}=\simeq 2.1$) the approximate solution overestimates the magnetization amplitudes.}
\end{figure}

Exact numerical results for the oscillation amplitude $\Delta M_{\textrm{AC}}$ corresponding to an anticrossing at $\nu=4$ versus azimuthal angle $\phi$ are shown as symbols interconnected by solid lines in figure~\ref{Fig6}.
 We consider a $7$~nm wide InSb QW with $n_{\textrm{s}}=2.5\times 10^{15}$~m$^{-2}$. The calculations are performed for $m^*=0.014m_e$, $g^*=-51$ and $\beta_{\textrm{D}}=3.22\times 10^{-11}$~eVm taken from \cite{Destefani2005}. $\alpha_{\textrm{R}}=\Delta_{\textrm{R}}/2k_{\textrm{F}}$ is varied as shown in the legend. These parameters lead to $\theta_{\textrm{c}}=69.08^{\circ}$. We find that the presence of, both, R-SOI and D-SOI provokes a pronounced anisotropy of $\Delta M_{\textrm{AC}}$ with respect to the direction of the in-plane magnetic field component. The anisotropic $\Delta M_{\textrm{AC}}$ contains information on the absolute values of $\alpha_{\textrm{R}}$ and $\beta_{\textrm{D}}$ as well as their relative sign as will be detailed later. Dashed lines in figure~\ref{Fig6} denote the approximate analytical results. Again, the analytical approximation is valid in the regime $\Delta M_{\textrm{AC}} B_{\perp} \ll \hbar \omega_{\textrm{c}}$. From the first order approximation, one gets \footnote{This equation holds for negative sign of $g^*$. For positive $g^*$, $\Delta_{\textrm{R}}^2$ and $\Delta_{\textrm{D}}^2$ need to be exchanged.}
\begin{eqnarray}
& \Delta M_{\textrm{AC}} (\phi) \approx \frac{1}{B_{\perp}}  \nonumber \\
& \times  \sqrt{ \Delta_{\textrm{R}}^2 \sin^4 \frac{\theta_{\textrm{c}}}{2} +\Delta_{\textrm{D}}^2 \cos^4 \frac{\theta_{\textrm{c}}}{2}+\frac{\Delta_{\textrm{R}} \Delta_{\textrm{D}}}{2} \sin^2 \theta_{\textrm{c}} \sin 2 \phi } \mbox{ .}\label{eq:anisotropic-gap}
\end{eqnarray} For the discussion we have chosen filling factor $\nu=4$, which occurs at $B_{\perp} \simeq 2.59$~T leading to $B\simeq 7.25$~T at the coincidence angle. This value is within experimental reach \cite{Wilde2008}. With the R-SOI strength approaching the D-SOI value from below, a pronounced anisotropy of the oscillation amplitude develops. The strength of the anisotropy depends on $\alpha_{\textrm{R}}/ \beta_{\textrm{D}}=\Delta_{\textrm{R}}/\Delta_{\textrm{D}}$. The minimum amplitude occurs at $\phi=-45^{\circ}$ and the maximum amplitude at $\phi=45^{\circ}$. The orientation of the double-loop figure depends on the relative sign of $\alpha_{\textrm{R}}$ and $\beta_{\textrm{D}}$ (bottom right inset of figure~\ref{Fig6}): For $\alpha_{\textrm{R}}/\beta_{\textrm{D}}=-2.1$, the position of minima and maxima occur at $\phi=45^{\circ}$ and $\phi=-45^{\circ}$, respectively. We note that the definition of the sign of $\beta_{\textrm{D}}$ varies in the literature \cite{Das1990,Fal'ko1992,Schliemann2003,Ganichev2004,Wang2005,Olendski2008}. The isotropic oscillation amplitude expected for the InGaAs/InP structure with $\Delta_{\textrm{D}}=0$ is shown in the top right inset for comparison. The case $|\Delta_{\textrm{R}}|=|\Delta_{\textrm{D}}|=\Delta$ is special due to its relevance for spintronics \cite{Schliemann2003} and because beating patterns do not exist in perpendicular magnetic fields. Here, (\ref{eq:anisotropic-gap}) reduces to $\Delta M_{\textrm{AC}} \approx \frac{\Delta}{B_{\perp}} \sqrt{1+\frac{1}{2} \sin^2 \theta_{\textrm{c}}(\sin 2 \phi \mp 1})$ and the value of $\Delta$ and the relative sign of the contributions can be extracted. The exact calculation and the analytical approximation match well as shown in figure~\ref{Fig6} as solid interconnected squares and dashed line, respectively.

\begin{figure}
\centering
\includegraphics{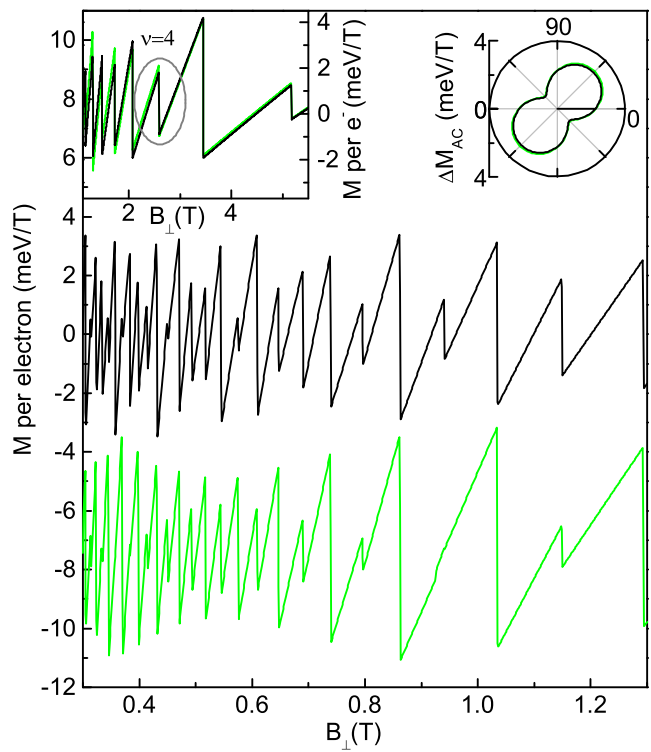}
\caption{Beating pattern in $M$ for InSb QW parameters at the coincidence angle $\theta_{\textrm{c}}=69.08^{\circ}$ and $\phi=45^{\circ}$. The curves for $\Delta_{\textrm{R}}=\Delta_{\textrm{D}}=8.07$~meV (dark) and $\Delta_{\textrm{R}}=17.03$~meV,  $\Delta_{\textrm{D}}=3.82$~meV (light) differ in the low field regime, where beating patterns are present. The curves are offset for clarity. In large $B_{\perp}$ the situation is different (upper left inset). Here, the magnetization traces are almost identical, since inside an anticrossing region (circle) the amplitudes are the same to first order as given by (\ref{eq:2-solutions}). Outside the anticrossing regions the spin splitting is governed by the (identical) Zeeman energy. Upper right inset: $\Delta M_{\textrm{AC}}(\phi)$ for $\nu=4$ for both parameter pairs. In this exact calculation, the degeneracy implied by the analytical approximation is lifted. \label{Fig7} }
\end{figure}

Starting values $\Delta_{\textrm{R}}$ and $\Delta_{\textrm{D}}$ for the exact modeling can be extracted from the maximal and minimal oscillation amplitudes in the experimental data occurring at $\phi=\pm \pi/4$ using the analytical approximation
\begin{equation}
\Delta M_{\textrm{AC}}^{\pm} (\phi=\pm \pi/4) \approx \frac{ \left| \Delta_{\textrm{R}} \sin^2 \frac{\theta_{\textrm{c}}}{2} \pm \Delta_{\textrm{D}} \cos^2 \frac{\theta_{\textrm{c}}}{2} \right|}{B_{\perp}} \mbox{ .} \label{eq:minmaxgap}
\end{equation} Here, we assume $\Delta_{\textrm{R}}/\Delta_{\textrm{D}}>0$ and $g^*<0$. For $\Delta_{\textrm{R}}/\Delta_{\textrm{D}}<0$ (\ref{eq:minmaxgap}) applies to $\phi=\mp \pi/4$. Note that for $\phi=-\pi/4$ we get $\Delta M_{\textrm{AC}}=0$ for $\alpha_{\textrm{R}}/\beta_{\textrm{D}}=1/\tan^2 \theta_{\textrm{c}}/2$, i.e., the anticrossing gap vanishes to first order for this specific in-plane field direction. For the InSb quantum well this condition is matched for $\alpha_{\textrm{R}}/\beta_{\textrm{D}} \simeq 2.1$.
Equation (\ref{eq:minmaxgap}) yields in general four solutions for $\Delta_{\textrm{R}},\Delta_{\textrm{D}}$. Acknowledging that we can only determine the relative sign of $\Delta_{\textrm{R}}$ and $\Delta_{\textrm{D}}$ but not the absolute sign of both we end up with two solutions
\begin{equation}
\Delta_{\textrm{R}} \approx \frac{(\Delta M^+\pm \Delta M^-)B_{\perp}}{2\sin^2 \frac{\theta_{\textrm{c}}}{2}} \mbox{ , } \Delta_{\textrm{D}} \approx \frac{(\Delta M^+\mp \Delta M^-)B_{\perp}}{2\cos^2 \frac{\theta_{\textrm{c}}}{2}} \mbox{ .} \label{eq:2-solutions}
\end{equation} These reduce to one for $\Delta_{\textrm{R}}/\Delta_{\textrm{D}}=1/\tan^2 (\theta_{\textrm{c}}/2)$, since here $\Delta M^-=0$. We point out that for the general case of two nondegenerate solutions, the correct solution is obtained by comparing the numerically calculated magnetization with the experiment: the two solutions are distinguished by the behavior of $M$ in the low-field regime where the beating patterns occur. To substantiate this, we show numerically calculated magnetization traces in figure~\ref{Fig7} for the case $\Delta_{\textrm{R}}=\Delta_{\textrm{D}}=8.07$~meV and the corresponding second solution of (\ref{eq:2-solutions}) given by $\Delta_{\textrm{R}}=17.03$~meV and $\Delta_{\textrm{D}}=3.82$~meV.
 While the behavior of $M$ at large $B_{\perp}$ (left inset of figure~\ref{Fig7}) is similar for both cases, the beating patterns at low fields differ (main graph). Thus, the exact diagonalization allows us to distinguish the two cases corresponding to the two solutions of (\ref{eq:2-solutions}) and identify $\alpha_{\textrm{R}}$, $\beta_{\textrm{D}}$ and their relative sign. As can be seen in the right inset at $\phi=45^{\circ}$ (arrow), the full numerical treatment in general lifts the degeneracy of the two solutions for $\Delta M_{\textrm{AC}}(\phi)$ present in the approximation of (\ref{eq:2-solutions}). We note here, that the same procedure applies when only one term of $\Delta_{\textrm{R}}$ an $\Delta_{\textrm{D}}$ is finite, and it is not known \emph{a priori} which on it is (unlike the situation considered in section~\ref{sec:Dominant_SOI_term}, where we considered R-SOI). In such a case $\Delta M_{\textrm{AC}}(\phi)=\Delta M$ is isotropic, and the two possible solutions are $\Delta_{\textrm{R}} \approx \Delta M B_{\perp}/ \sin^2 \frac{\theta_{\textrm{c}}}{2}$ and $\Delta_{\textrm{D}} \approx \Delta MB_{\perp}/\cos^2 \frac{\theta_{\textrm{c}}}{2}$. Again, the applicable solution can be found from the low-field behavior of $M$, since the beating patterns for the two solutions differ.

The analysis of the anisotropy of $\Delta M_{\textrm{AC}}$ based on (\ref{eq:2-solutions}) can be used as a starting point for the exact diagonalization. The full numerical treatment then provides the definite values of $\alpha_{\textrm{R}}$ and $\beta_{\textrm{D}}$ by one-to-one comparison with the experimental data over a broad field regime.

\subsection{2DESs in wide quantum wells: $k^3$ D-SOI}\label{sec:cubicDSOI}

\begin{figure}
\centering
\includegraphics{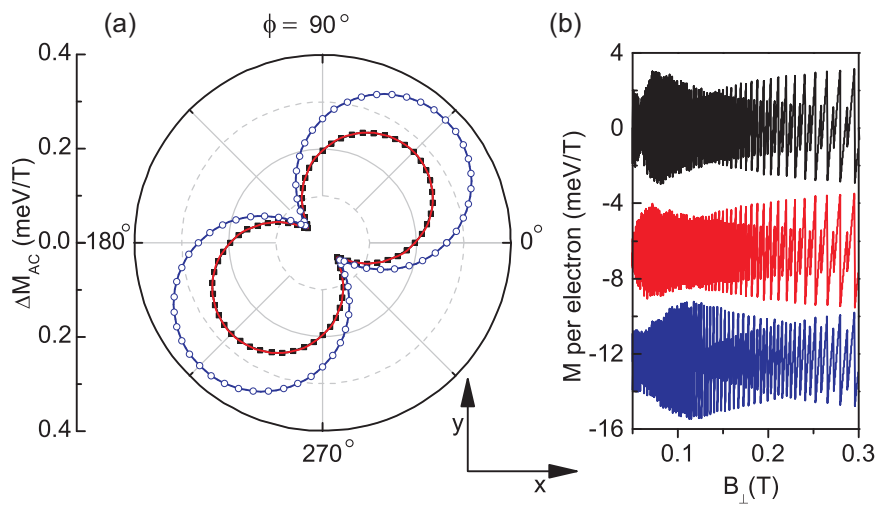}
\caption{\label{Fig8} (a) Polar plot of $\Delta M_{\textrm{AC},\nu=4}$ vs azimuthal angle $\phi$. Calculations are performed for $\alpha_{\textrm{R}}=5.1\times 10^{-12}$~eVm, $\beta_{\textrm{D}}=3.2\times 10^{-12}$~eVm and $\gamma_{\textrm{D}}=4.5\times 10^{-28}$~eVm$^3$ given in \cite{Gilbertson2008} for a $30$~nm wide In$_{0.85}$Al$_{0.15}$Sb/InSb/In$_{0.9}$Al$_{0.1}$Sb asymmetric quantum well. Solid squares interconnected by lines represent the values including the full Hamiltonian (\ref{eq:H-cubic}), i.e., including all $k^3$ D-SOI terms. Open circles interconnected by lines denote results neglecting $k^3$ D-SOI. The curves differ in the absolute values, highlighting the importance of $k^3$ D-SOI in wide quantum wells. The solid red line shows the result of a calculation where only the $k^3$ D-SOI matrix element is included that couples neighboring Landau levels as discussed in the main text. (b) Low field beating patterns in $M$ for $\theta =\theta_{\textrm{c}}$ and $\phi=45^{\circ}$ including $k^3$ D-SOI (upper curve) and neglecting $k^3$ D-SOI (lower curve). The curves are shifted in vertical direction for clarity. }.
\end{figure}

We have focussed on the limit of narrow quantum wells and low $n_{\textrm{s}}$ up to now. This allowed us to neglect the Dresselhaus terms that are cubic in the wave vector $k$. Such conditions are met for many realistic systems and we showed that in this limit it is in particular possible to use the analytical approximation (\ref{eq:anisotropic-gap}) for the analysis of experimental data. With the numerical model at hand it is however instructive to consider the impact of the $k^3$ D-SOI in the limit of wide quantum wells.
In the following we thus consider the Hamiltonian \cite{Das1990,Gilbertson2008}
\begin{equation}
H=H_0+\frac{\alpha_{\textrm{R}}}{\hbar} \left( \sigma_x \pi_y - \sigma_y \pi_x \right)+ \frac{\beta_{\textrm{D}}}{\hbar} \left( \sigma_x \pi_x - \sigma_y \pi_y \right)+ \frac{\gamma_{\textrm{D}}}{\hbar} \left( \sigma_y \pi_x^2 \pi_y - \sigma_x \pi_x \pi_y^2 \right)\mbox{ .} \label{eq:H-cubic}
\end{equation}
For lack of experimental values in the literature, we consider values $\alpha_{\textrm{R}}=5.1\times 10^{-12}$~eVm, $\beta_{\textrm{D}}=3.2\times 10^{-12}$~eVm and $\gamma_{\textrm{D}}=4.5\times 10^{-28}$~eVm$^3$ calculated in the extensive work of Gilbertson \emph{et al.} \cite{Gilbertson2008} for a $30$~nm wide In$_{0.85}$Al$_{0.15}$Sb/InSb/In$_{0.9}$Al$_{0.1}$Sb asymmetric quantum well with $n_{\textrm{s}}=2.5\times 10^{15}$~m$^{-2}$. We show the results in figure~\ref{Fig8}.
Solid squares interconnected by lines represent the values including the full Hamiltonian (\ref{eq:H-cubic}), i.e., including all $k^3$ D-SOI terms. For comparison, we show the result where $k^3$ D-SOI is neglected as open circles interconnected by lines. The difference in the absolute values shows that the $k^3$ D-SOI is important in such wide quantum wells, in contrast to the systems considered in the previous sections. However, the angular dependence is qualitatively the same. This raises the question, whether one can determine the $k^3$ D-SOI contribution from experimental data on $\Delta M_{\textrm{AC}}$ and if analytical approximations similar to (\ref{eq:anisotropic-gap}) can be found. To answer this, it is more instructive to consider the matrix elements of the problem as given in the appendix of \cite{Das1990}, i.e. without rotation of the spin quantization axis. This minimizes the number of entries in the matrix that are relevant for the discussion. In brief, the last term in (\ref{eq:H-cubic}) leads to two non-vanishing matrix elements. One couples different spin levels that differ in the Landau index by $3$, i.e., it couples levels that are far apart in energy in high fields. A calculation of the anticrossing gap neglecting this matrix element [solid red line in figure~\ref{Fig8}] is virtually identical to the calculation for the full Hamiltonian [solid squares].

The other matrix element couples exactly the same levels as the only non-vanishing $k$-linear Dresselhaus matrix element, i.e., different spin levels of neighboring Landau levels. It has opposite sign compared to the $k$-linear element, leading to a decreased value of this entry. Fitting (\ref{eq:anisotropic-gap}) to experimental data thus yields useful information also in the limit of wide quantum wells, when $\Delta_{\textrm{D}}^{'}=2k_{\textrm{F}}(\beta_{\textrm{D}}-\gamma_{\textrm{D}} k_{\textrm{F}}^2/4)$ is substituted for $\Delta_{\textrm{D}}$. Whether the $k^3$ D-SOI can be neglected in the analysis of experimental data $\Delta M_{\textrm{AC}}$ can again be determined experimentally from the different beating patterns in low magnetic fields. This is demonstrated in figure~\ref{Fig8}~(b), where we show the low field beating patterns at $\theta_{\textrm{c}}$ predicted for the full Hamiltonian including the $k^3$ D-SOI (upper curve) and neglecting the $k^3$ D-SOI (lower curve). Details of the anomalous beatings that occur in $M$ due to (\ref{eq:H-cubic}) in perpendicular magnetic fields have been discussed in \cite{eSilva1994}. Information about the relative strength of $\beta_{\textrm{D}}$ and $\gamma_{\textrm{D}}$ could be obtained by variation of $k_{\textrm{F}}=\sqrt{ 2\pi n_{\textrm{s}}}$ via the carrier density.

\section{Discussion}\label{sec:discussion}

In the following we discuss the observability of the predictions using state-of-the-art experimental techniques. The magnetization of single-layered 2DESs in semiconductor heterostructures is weak and has been resolved after optimizing custom designed magnetometers. These include torsion wire magnetometers \cite{Eisenstein1985b,Wiegers1997,Zhu2003,Schaapman2002}, a custom designed superconducting quantum interference device (SQUID) \cite{Meinel1997}, and micromechanical cantilever magnetometers \cite{Harris1999,Schwarz2002,Wilde2008}. A field modulation technique \cite{Anissimova2006} was also applied to measure the imaginary part of the AC current between a gate electrode and the 2DES. Commercially available magnetometers are typically not sensitive enough \cite{Wilde2010}. SQUID magnetometry and the field modulation technique might be especially useful for measuring in weak perpendicular magnetic fields. These techniques have not yet been demonstrated in tilted magnetic fields as required for the predicted magnetization signals. For torsion wire balance and cantilever magnetometers, the magnetic torque signal increases with the total magnetic field. Recently, micromechanical cantilever magnetometry has been performed in large fields where the tilt angle was varied systematically \cite{Wilde2005,Wilde2009,Windisch2009,Rupprecht2013}. A resolution of $M$ per electron better than $0.001$~meV/T has been demonstrated for conditions \cite{Wilde2010} that would be required for the proposed investigations. Cantilever magnetometers are thus particularly suited for experiments addressing the anisotropic $M$.

In general, the magnetization experiment needs to be performed as a function of both the out-of-plane and in-plane field angles $\theta$ and $\phi$, respectively. For the case where it is known \emph{a priori} that either R-SOI or D-SOI is absent, the experiment requires rotation about $\theta$ only. To be specific, for the InGaAs/InP quantum well parameters the Zeeman-induced artificial level degeneracy (without SOI) is expected to occur at $\theta_{\textrm{c}} \simeq 85.28^{\circ}$. The purely SOI-induced dHvA oscillation at $\nu=4$ highlighted in figure~\ref{Fig3} occurs at a perpendicular magnetic field $B_{\perp}\simeq 2.26$~T. This corresponds to a total magnetic field of $B\simeq 27.6$~T. Cantilever magnetometry in tilted fields up to $B=33$~T has been demonstrated in \cite{Wilde2005}. After measuring $\Delta M_{\textrm{AC}}(\theta)$ and determining its minimum value, the Rashba parameter $\alpha_{\textrm{R}}$ is extracted from modeling of the data.

When both R-SOI and D-SOI play a role, the experimental procedure is as follows: First, the coincidence angle $\theta_{\textrm{c}}$ is determined from the measured minimum of $\Delta M_{\textrm{AC}}(\theta)$ for a given $\phi$. Second, $\theta=\theta_{\textrm{c}}$ is fixed and the field is rotated in the plane about $\phi$. This procedure automatically accounts for a $g$-factor anisotropy between out-of-plane and in-plane directions \cite{Kalevich1992} in that the experimentally determined critical angle $\theta_{\textrm{c}}$ corresponds to the $g$-factor for that specific field direction. In case of an in-plane anisotropy of the $g$-factor \cite{Kalevich1993}, the coincidence angle $\theta_{\textrm{c}}$ becomes a function of $\phi$ and should thus be determined for each $\phi$. Modeling the measured anisotropic behaviour of $\Delta M_{\textrm{AC}}$ using the numerical model introduced above yields the absolute values of $\alpha_{\textrm{R}}$ and $\beta_{\textrm{D}}$ and their relative sign. For the InSb based quantum well, the coincidence condition amounts to $\theta_{\textrm{c}} \simeq 69.08^{\circ}$. At this angle, the filling factor $\nu=4$ discussed above occurs at $B\simeq 7.6$~T, which can be achieved in a conventional superconducting solenoid. An experiment in a doubly tilted configuration needs to be performed. This is expected to be experimentally challenging. However, measurements at only two in-plane angles $\phi=\pm \pi/4$ are in principle sufficient to determine both SOI constants and their relative sign by comparison to the model calculation.

We have focussed on the case of an ideal 2DES at zero temperature $T$ for the sake of clarity. Considering a finite temperature $T$ and disorder broadening of the LLs in the calculations is straightforward as demonstrated in figure~\ref{Fig4}. This way a one-to-one correspondence between experimental and theoretical data becomes possible for a real 2DES.

\section{Conclusions}\label{sec:conclusions}

Considering recent advances in magnetometry on 2DESs we have presented a detailed numerical analysis of the quantum oscillatory magnetization when Rashba and Dresselhaus SOI are relevant. The formalism predicts that R-SOI and D-SOI constants $\alpha_{\textrm{R}}$ and $\beta_{\textrm{D}}$, respectively, can be extracted from the de Haas-van Alphen effect if measured in tilted magnetic fields. For the evaluation, the anticrossing of energy levels is found to be of particular importance. In the case of one dominant SOI term, the magnetization oscillation amplitude at the anticrossing is a direct measure of the SOI strength. In the general case of $\alpha_{\textrm{R}} \neq 0$ and $\beta_{\textrm{D}} \neq 0$, the oscillation amplitude exhibits a pronounced anisotropy with respect to the direction of the in-plane magnetic field component. Using realistic sample parameters we argue that state-of-the-art torque magnetometry techniques allow one to study experimentally the predicted anisotropy of $M$. Via numerical modeling $\alpha_{\textrm{R}}$ and $\beta_{\textrm{D}}$ as well as their relative sign are obtained. Experiments addressing $M$ lift ambiguities and circumvent model assumptions that exist with the analysis of beating patterns in magnetotransport data and the investigation of weak-antilocalization features, respectively. Magnetization measurements thus allow one to further promote the understanding of SOI in low-dimensional electron systems.

\ack

We thank E.~I. Rashba for valuable discussions and gratefully acknowledge financial support by the DFG via SPP1285, grant no. GR1640/3.

\appendix

\section*{References}

\bibliography{biblio}


\end{document}